# Very-Large-Scale Reconfigurable Intelligent Surfaces for Dynamic Control of Terahertz and Millimetre Waves

Yury Malevich[1,2], Said Ergoktas[1,2], Gokhan Bakan[1,2], Pietro Stainer[1,2]

Coskun Kocabas[1,2,3,*]

[1]Department of Materials, University of Manchester, Manchester, M13 9PL, UK

[2]National Graphene Institute, University of Manchester, Manchester, M13 9PL, UK

[3]Henry Royce Institute for Advanced Materials, University of Manchester, Manchester M13 9PL, UK

*Corresponding authors E-mail: coskun.kocabas@manchester.ac.uk

**Unlocking the potential of terahertz (THz) and millimetre (mm) waves for next generation communications[1–4] and imaging[3,5,6] applications requires reconfigurable intelligent surfaces (RIS) with programmable elements that can manipulate the waves in real-time[7]. Realization of this technology has been hindered by the lack of efficient THz electro-optical materials and THz semiconductor platform. Here, by merging graphene-based THz modulators and the thin-film transistor (TFT) technology, we demonstrate mega-scale spatial light modulator arrays with individually addressable subwavelength pixels. We demonstrate electronically programmable patterns of intensity and phase of THz light over a large area with unprecedent levels of uniformity and reproducibility. To highlight the potential of these devices, we demonstrate a single pixel mm-wave camera capable of imaging metallic objects. We anticipate that these results will provide realistic pathways to structure THz waves for applications in non-invasive THz imaging and next generation THz communications.**

Due to material and technological challenges, the terahertz (0.3 to 3 THz) and sub-terahertz (90–300 GHz) region remains a relatively underutilized portion of the electromagnetic spectrum, despite its wide range of promising applications in communication, sensing, and imaging technologies[5,8–11]. The advancement of future THz technologies necessitates devices capable of generating spatio-temporal patterns of THz light. Studies on THz optoelectronics have led to several successful demonstrations using active metamaterials[12–14], liquid crystals[15–17], phase-change materials[18–20], plasmonic modulators[21] and photoinduced electrons[22]. Although these results are impressive at the single-device level, they encounter challenges in large-scale system-level integration, critical for unlocking new possibilities for scalable THz technologies. A promising alternative involves controlling high-mobility charges on graphene[23–28] via electrostatic gating in a transistor configuration. These devices leverage the tuneable Drude-like metallic response of graphene at THz frequencies. However, the developmental stage of these devices is still nascent compared to those operating in the visible spectrum. Overcoming these developmental hurdles would bridge existing technological gaps for emerging THz technologies.

In this work, we introduce an integration scheme that incorporates large arrays of graphene-based THz modulators, allowing programmable control over reflection and transmission across the wide THz and mm-wave spectrum. Figure 1a depicts the structure of a pixel, highlighting the key functional layers: graphene, electrolyte, and back-plane electronics. In this architecture, the THz active layer consists of bilayer graphene, produced by chemical vapor deposition on copper foils, and subsequently transferred onto a thin polymer sheet (approximately 70 µm PET, see supporting materials Figure S1). A 5 µm-thick, porous membrane infused with an ionic liquid electrolyte is laminated between the graphene layer and the partially transparent Indium Tin Oxide (ITO) pixel electrode. In this device, the thickness of the electrolyte layer is critical to minimize electrostatic crosstalk between pixels ( supporting materials Figure S2) enabling area selective charge accumulation on continuous graphene layer. The charge accumulated on the pixel capacitor (formed between the middle electrode and ground plane, as shown in Figure 1a) dictates the local hole/electron density on the graphene layer through electrolyte gating, thereby modulating THz reflection and transmission. We applied an additional negative bias voltage ($V_G$) on the graphene layer to offset the shift of the Dirac point due to unintentional doping during the fabrication process.

The back-plane electronics, illustrated in Figures 1b and 1c, utilize active-matrix technology to regulate the local voltage of individually addressable sub-wavelength pixels within a dense array of 640 rows and 480 columns covering $12 \times 9$ cm$^2$ area. The inset in Figure 1c displays an enlarged image of a pixel, which contains an n-type amorphous-Si TFT with a double-back-gate configuration (Figure 1d). The TFT, pixel capacitor, and voltage lines are covered by a 2 µm thick polymer layer. The top ITO layer maintains direct contact with the electrolyte and is connected to the pixel electrode through a metalized via.

Figure 1e presents the output and transfer characteristics of the n-type a-Si TFT. The charge on the pixel capacitor is controlled by applying positive voltage pulses on the gate while switching between high ($V_{DH}$) and low drain ($V_{DL}$) voltages with scan rate of 200 Hz. Due to the current saturation at positive drain voltages, the charging and discharging currents display a significant difference, which is compensated by the duration of applied gate pulses. The drain and gate voltages are scanned over the modulator array by a Chip-on-Glass display driver capable of generating a binary voltage pattern for $V_{DH}$ and $V_{DL}$. The effective pixel voltage is then determined by the width of the gate pulse (varied from 30 to 208 µs), as depicted in Figure 1f. To visualize the generated voltage pattern at single pixel level, we recorded an electron microscope image of the operating TFT array using a secondary electron detector. The contrast in Figure 1g results from the voltage difference between the pixels. Pixels with negative voltages generate a greater number of secondary electrons, thereby enhancing the contrast significantly.

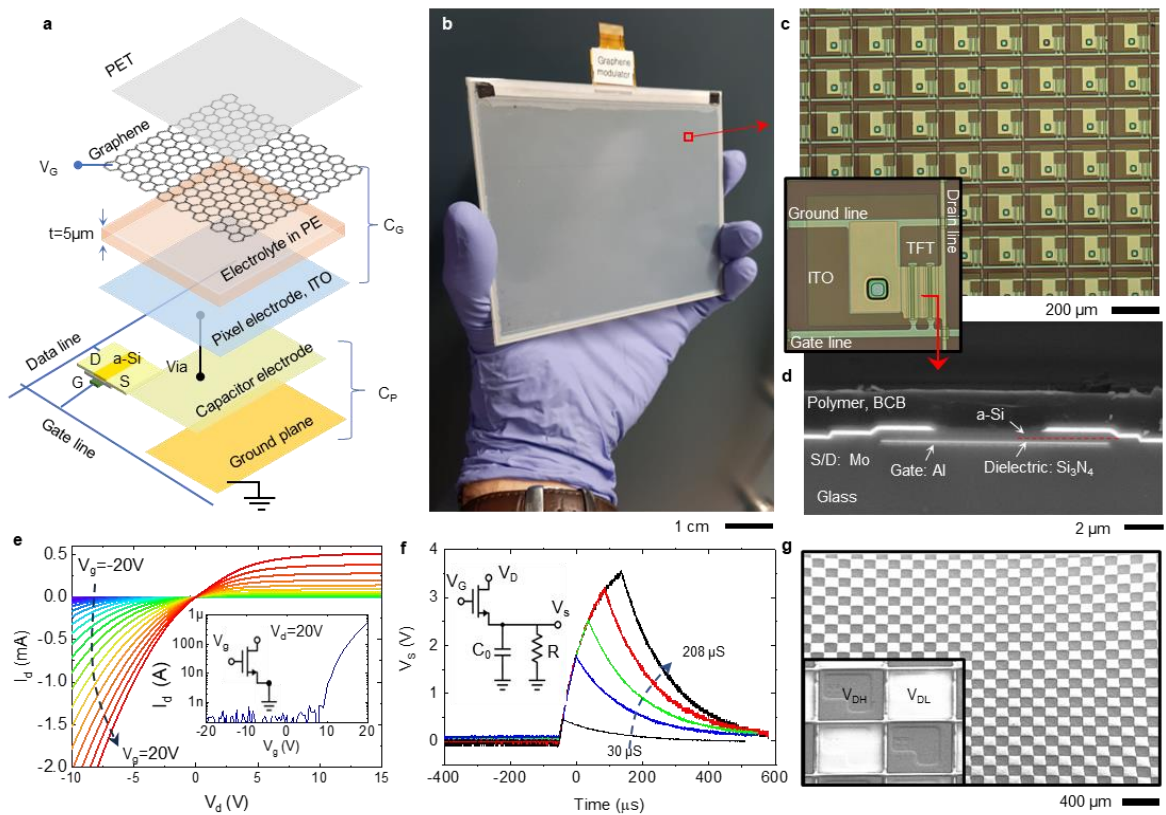

**Figure 1. Reconfigurable THz Surface: a**, Schematic of the pixel structure, comprising laminated layers including a graphene top electrode, an electrolyte layer, and a back pixel electrode. Common voltage, $V_G$, is applied to the graphene layer to compensate unintentional doping, thereby enabling the device to operate near the Dirac point. The charge on the capacitor is regulated by the duration of gate voltage pulses applied to the gate line. **b**, Photograph of the fabricated device consisting of an active-matrix array of 640×480 pixels. A binary voltage

pattern ($V_{DH}$, $V_{DL}$) is produced by a Chip-on-Glass display driver controlled by an external microcontroller. **c**, Photograph of the TFT back panel, displaying the sub-wavelength pixels, each consisting of a top ITO electrode connected to the pixel capacitor with a via through the polymer layer. The inset illustrates the unit cell with the size of 185 µm. A double back-gated a-Si thin-film transistor, with a channel length of 5µm and a width of 75 µm, governs the charge/voltage on the pixel capacitor, which is formed between the middle capacitor electrode and the ground plane. **d**, Scanning electron micrograph revealing the cross-section of the a-Si thin-film transistor. **e**, Representative output ($I_d$ vs $V_d$) and transfer ($I_d$ vs $V_g$) characteristics of the n-type a-Si TFT. **f,** Analysis of the pixel voltage's dependence on the width of the applied gate pulses. The inset presents the circuit model of the pixel, incorporating the probe resistance (1 MΩ) and capacitance (16pF), resulting in a 1.5 ms charging time. **g**, Scanning electron microscope image of the surface, with charging contrast induced by the alternating voltages of $V_{DH}$ and $V_{DL}$ on the pixels.

Our reconfigurable surfaces operate in both transmission and reflection modes, offering dynamic control over terahertz (THz) and millimetre waves. First, we analysed the transmission mode by measuring the programmable spatio-temporal patterns of THz transmission using a sensor array (Terasense, 64×64 array with 1.5 mm pixel pitch) and illumination with a collimated 100 GHz beam generated by an IMPATT diode (Figure 2a). Figure 2b shows representative images of letters N, G, I (the initials of the National Graphene Institute) and different transmission patterns (see Supporting Video1 and 2). The spatial resolution of these recorded images is limited by pixel size of the sensor array and the wavelength of the source (λ~3mm).

To optimise THz transmission modulation, we fabricated modulators with single, double and three layers of graphene. Figure 2c shows a comparison of the THz modulation spectrum derived from various devices using a time-domain THz spectrometer. The device with the double layer of graphene marked a significant enhancement over the one with a single layer; however, the introduction of a third layer resulted in considerable insertion loss without any improvement in modulation efficiency. This diminishing performance is presumably attributable to the inefficient gating of the subsequent graphene layers.

The drain voltage and the duration of the gate pulse are the two main control parameters of the modulator array. Figure 2d illustrates the dependence of the THz modulation (@ 1 THz) on the drain voltage. To quantify the charging time of the pixel, we applied pulses of VDH and VDL and monitored the transmission with a fast zero-bias Schottky diode attached to a horn antenna (Figure 2e). We observed that the charging and discharging time can be reduced down to 1ms with short gate pulses of 30μs (see the inset in Figure 2e).

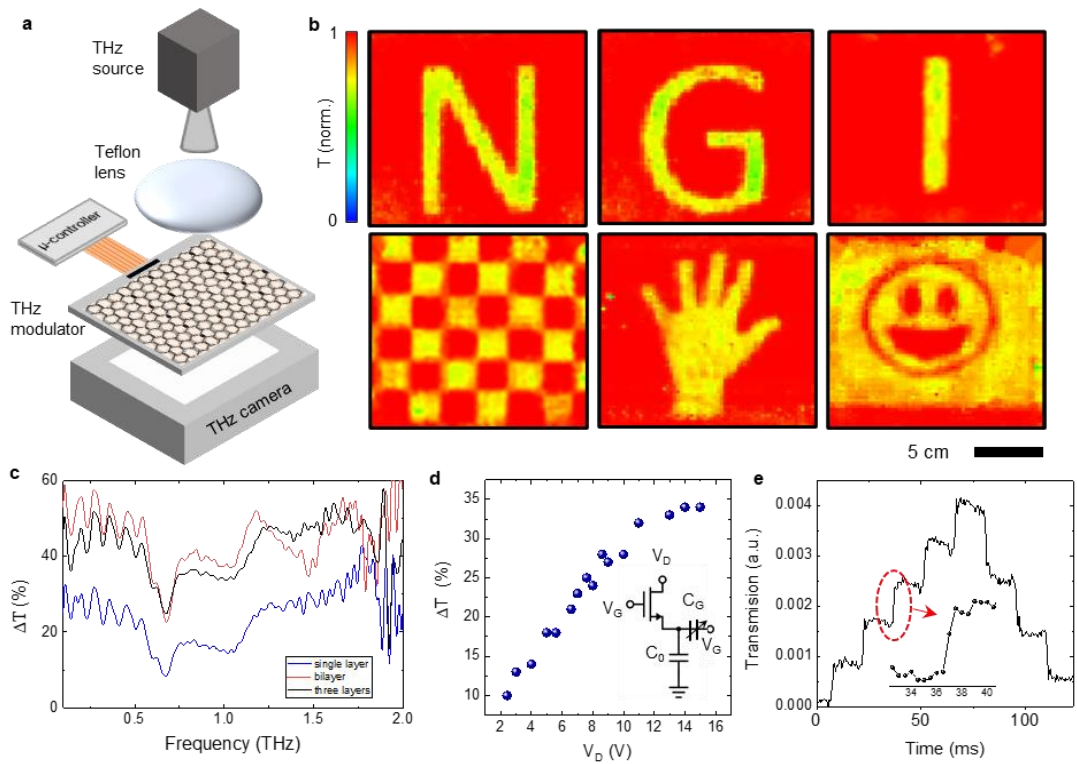

**Figure 2: Programmable THz transmission patterns**: **a**, Schematic of the experimental setup used for measuring transmission pattern using a 64x64 pixels THz camera. **b**, Representative images generated by the THz modulator array. **c**, Spectrum of the THz modulation for 3 different devices with single, double and three layers of graphene **d**, Voltage dependence of the modulation. The inset shows the circuit model where $C_0$ and $C_G$ represents the pixel capacitance and the voltage dependent capacitance of the electrical double layer, respectively. **e**, Variation of the transmission at 0.1 THz after applying consecutive charging and discharging pulses. The inset shows the transition region with switching time around 1ms.

Next, we investigate the reconfigurable THz reflectivity. To obtain large phase and intensity modulation, we have designed the device to operate around a reflection singularity which emerges from the interference of multiple reflections from the top polymer layer ($t_1$=70 μm), graphene and TFT array[29]. In this configuration, graphene behaves as a tuneable reflectivity mirror controlling the ratio of the interfering components (Figure 3a). This multilayer structure behaves as a tuneable coupled cavities. To observe the resonance around at THz frequencies, we increased the thickness of the electrolyte layer to $t_2=25$ µm. This configuration enables continuous tunability of the resonance frequency between $f_{max} \sim \frac{1}{t_1}$ and $f_{min} \sim \frac{1}{t_1+t_2}$. When graphene is reflective (@ $V_{DH}$), the resonance frequency is determined by the top polymer coating ($t_1 \sim 70$ µm), however around the Dirac point (@ $V_{DL}$), graphene layer is transparent therefore the resonance is mainly defined by the total thickness ($t_1+t_2 \sim 95$ µm). We mounted the device on a motorised xy-stage and measured the time-domain THz reflection spectrum using a 4-mirror reflection head. Figure 3b and c shows the variation of reflection amplitude and phase spectrum as the pixel voltage changes from $V_{DL}=-2.4$V to $V_{DH}=10$V. During this process, the resonance frequency was observed to vary from 1.6 to 1.9 THz. In addition to the continuous tunability, the reflection spectrum goes through a perfect absorption indicating a topological switching clearly evidenced by a step-like jump in the phase spectrum (Figure 3c). Besides the tuneable phase shift, this phase jump provides a modulation up to $2\pi$. To characterize reconfigurable THz patterns, we generated binary images on the modulator array and performed a raster scan of the THz reflectivity to map these patterns. The spatial resolution of our THz spectrometer is approximately 600 µm, which is sufficient to resolve fine details in the THz reflectivity patterns. Figure 3d shows reflectivity map ( $\Delta R = R(V_{DH}) - R(V_{DL})$ ) at 1.5 THz of checkerboard pattern generated by a binary image on the modulator. It is also possible to generate a grayscale THz pattern by defining a supercell by grouping a 3x3 array of pixels with binary voltages (see supporting materials Figure S5). In this configuration, a 10-step grayscale value can be achieved by averaging the binary values of the pixels in the supercell. Figure 3e shows a reflection map of a fork diffraction grating obtained by superimposing linear diffraction grating pattern with a spiral phase structure. Besides its intricate structure, this pattern contains a dislocation with a topological charge, enabling diffracted beam with vortices, which can be used in various applications like beam shaping, and holography (see supporting materials Figure S6). To test the limits of our approach we also mapped a complex high-resolution pattern, a portrait of Queen Elizabeth II (Figure 3f) showing the grayscale THz reflectivity.

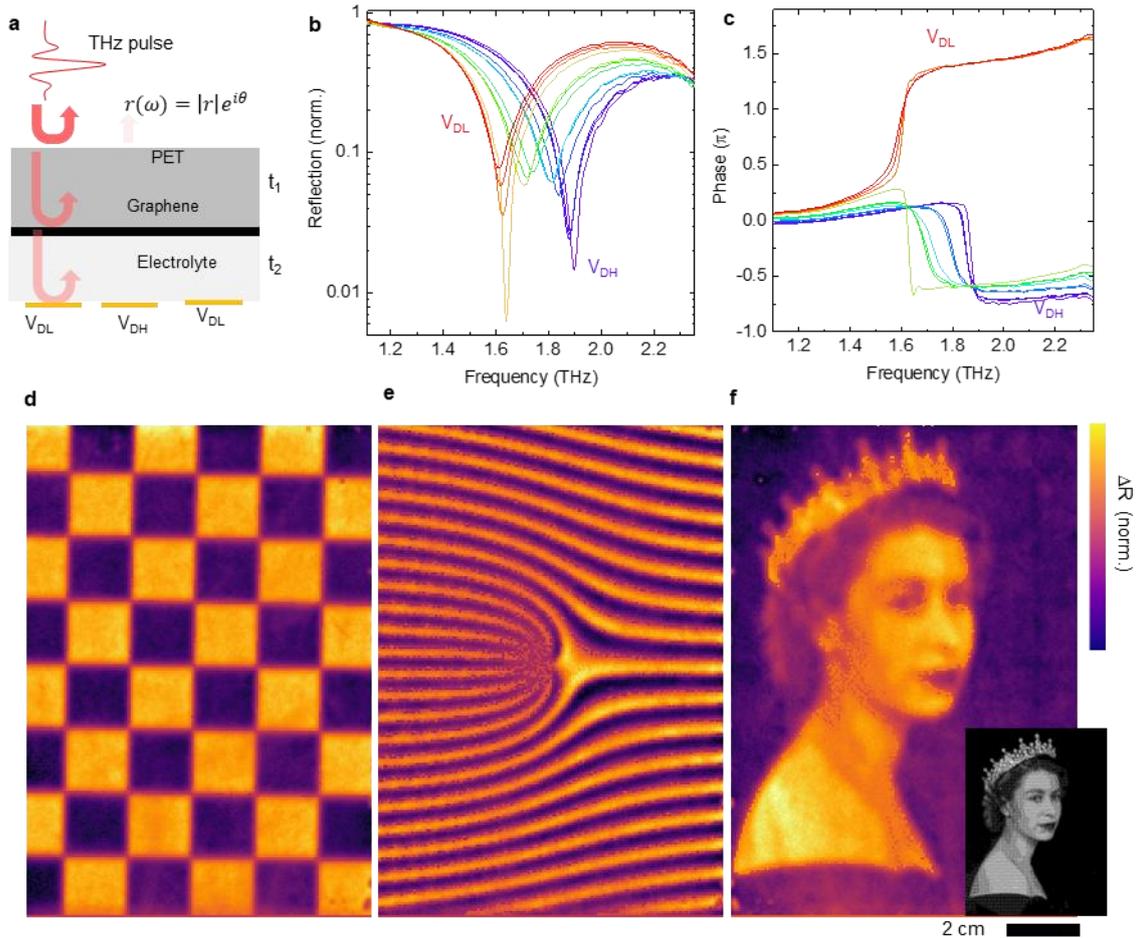

**Figure 3: Reconfigurable THz reflective surfaces**: **a**, Structure of the modulator showing the interfaces responsible for the multiple THz reflection resulting a tuneable resonance behaviour. **b, c** shows the variation of the intensity and phase spectrum as the pixel voltages switched between $V_{DL}$=-2.4V to $V_{DH}$=10V. The step like phase modulation is due to a topological phase transition enabled by the reflection singularity. **d,e,** and **f** show THz reflectivity map of a checkerboard, a grayscale fork diffraction grating and a grayscale portrait of Queen Elizabeth II. Inset shows the digitised binary image using 3x3 supercell.

To showcase a practical application of these large-scale modulators, we built a single pixel camera capable of imaging concealed metallic objects. Figure 4a shows the experimental setup used for the system. The modulator array is illuminated by a collimated beam of 100 GHz light. The total transmitted light is collected and measured with a pyroelectric sensor. A metallic object placed in a paper envelope is positioned in front of the modulator. The core working principle of the camera is based on compressed sensing algorithm[13,30–32]. This algorithm enables the reconstruction of the spatial information of the object (X, the vector representing the transmittance of the object) by measuring a series of structured transmission patterns generated by the modulator. We created a set of Hadamard transmission masks (M, matrix containing the mask set) on the modulator and recorded the transmitted intensity (I, intensity array). For each mask, we measured the differential modulation for by changing the pixel voltage between $V_{DL}$ and $V_{DH}$. This approach generated positive or negative values which is required to construct the orthogonal set of Hadamard masks to decode patterns as $X=M^{-1} I$. Figure 4b and c shows reconstructed 32×32 pixel images of a wrench and a razor blade. It is important to note that the resolution of the imaging system is limited by the wavelength of the source (λ=3mm) used for the illumination. Notably, the cutout width of razor blade, approximately 2mm, is visible in the reconstructed image. The graph on Figure 4d shows the convergence of image error function ($\chi^2$) with the mask number used in the reconstruction process.

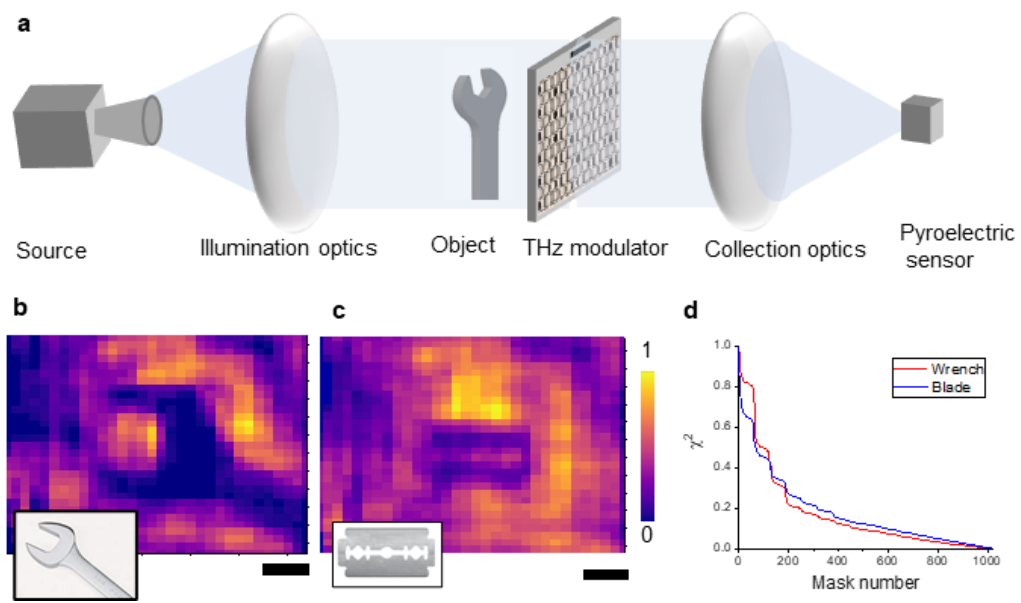

**Figure 4**: Single-pixel millimetre wave camera. **a**, Experimental setup used for the single pixel camera consisting of a mm-wave source (100 GHz), illumination and collection optics, reconfigurable modulator array and a single pyroelectric sensor. The metallic object is placed in front of the modulator. **b**, and **c** show the reconstructed images of a wrench and razor blade. The scale bars are 2 cm. **d**, shows the convergence of the error function as a function of mask number used in the reconstruction algorithm.

As a conclusion, by merging graphene modulators and the active-matrix TFT technology, we demonstrate large-area reconfigurable surfaces with sub-wavelength programmable elements that can manipulate THz- and mm-waves in real-time. These devices can create reconfigurable spatio-temporal patterns of THz light both in transmission and reflection modes. We demonstrate that the reflection performance of these devices can be significantly improved by operating around a reflection singularity which creates large phase modulation up to $2\pi$. Our results suggest that these surfaces can be used as a phased array or spatial light modulator to dynamically steer THz light or to create complex beams which is a critical capability required for next generation communication systems. Furthermore, the proof-of-concept demonstration of the single-pixel THz imaging system highlight potentials for THz imaging for applications in security, medical imaging, and industrial inspection, where non-invasive inspection methods are essential.

**Methods:**

**Fabrication of the Modulator**: We purchased A4-size single-layer graphene, synthesized via CVD on copper foil, from MCK Tech Co. Ltd. A 70 μm-thick Polyethylene terephthalate (PET) polymer layer was laminated onto the copper foil at 130°C. Following this step, the copper layer was etched away in a 0.1 M ammonium persulfate (APS) solution, and the remaining graphene on PET was rinsed with deionized water. The modulator array was created by laminating a 25- or 5-μm-thick porous polyethylene layer containing ionic liquid electrolyte (DEME TFSI) onto the TFT back plane. Electrical contacts were deposited at the corners of the graphene layer and connected to the voltage source on the modulator. The custom design TFT back plane electronic is obtained from a third-party display manufacturer.

**Time-Domain THz Spectroscopy**: We measured the intensity and phase of THz light using a time-domain THz spectrometer, TeraFlash by Toptica Photonics, capable of acquiring spectra at a rate of 16 spectra/s. This spectrometer facilitates the THz reflection characterization of the device across a spectral range up to 4 THz and with over a 90 dB dynamic range. Two InGaAs fiber-coupled antennas, along with the THz optical path, are housed within a 4-mirror reflection head that is continuously purged with dry nitrogen. This setup eliminates water absorption lines from the THz spectra during reflectivity measurements. The device is mounted on two perpendicular X and Y motorized stages (Thorlabs linear translational stages), allowing movement in the plane perpendicular to the THz beam, while keeping the beam focused on the modulator surface. A custom LabView code simultaneously controls the THz spectrometer, the stages and the THz modulator. Raster scanning is performed in 0.5 mm steps. The system moves the modulator and acquires the reflection spectrum, while the voltage on the modulator array alternates between VDH and VDL. We obtain the THz amplitude and phase spectrum using an FFT algorithm with appropriate zero padding to achieve a spectral resolution of 1 GHz.

**Imaging with THz Camera**: We assessed the modulator's performance and its capability to generate complex patterns using a Terasense camera (based on plasmon oscillations on GaAs high-mobility heterostructure) with a 64x64 array with a 1.5 mm pixel pitch. The modulator was illuminated with a collimated 100 GHz beam produced by an IMPATT diode, Terasense, 80mW output power. To minimize light scattering, the modulator was placed directly over the THz camera array.

**Single Pixel THz Camera**: We developed a THz imaging system utilizing a continuous-wave 100 GHz beam from an IMPATT diode, Terasense, 80mW output power. This beam, collimated by an 8″ parabolic mirror or Teflon lenses, passes through both the modulator and the object. It is then focused by another 8″ parabolic mirror or Teflon lens into a pyroelectric detector, Gentec-EO model. A Raspberry Pi 3 Model B served as the electronic controller, transmitting Hadamard patterns to the modulator. Signal reading from the pyroelectric detector and synchronization with the Raspberry Pi and THz spectrometer were managed by a LabView-based FPGA and microprocessor board, myRIO-1900 from National Instruments.

**Electromagnetic Simulations**: Using COMSOL-Multiphysics for electromagnetic simulation, we modelled the charge density on graphene. These simulations demonstrated that charge density on continuous graphene can be modulated at a single pixel level. The electrolyte layer's thickness and the pixel coverage are crucial to minimize crosstalk effects between pixels on the graphene layer.

**Acknowledgements**: This research is supported by Defence Science and Technology Laboratory (DSTLX-1000135951) and EPSRC EP/X027643/1 (ERC PoC grant).

**Competing interests**: The work is subject to a patent application by the University of Manchester.

**Author contributions**: Y.M. , M.S.E. G.B. and C.K. designed and fabricated the devices. Y.M. and M.S.E. built the setups and performed the experiments. G.B. developed the code for the TFT array and performed the electronical characterization of the TFTs. P.S helped for the experiments and sample preparation. Y.M., MSE, C.K. analysed the data and wrote the manuscript with feedback from the other authors. All authors discussed the results and contributed to the scientific interpretation as well as to the writing of the manuscript.